\documentclass[conference]{IEEEtran}
\IEEEoverridecommandlockouts
\usepackage{cite}
\usepackage{amsmath,amssymb}
\usepackage{algorithmic}
\usepackage{graphicx}
\usepackage{textcomp}
\usepackage{xcolor}
\def\BibTeX{{\rm B\kern-.05em{\sc i\kern-.025em b}\kern-.08em
    T\kern-.1667em\lower.7ex\hbox{E}\kern-.125emX}}
    
    \usepackage{euscript}
\usepackage{bbm}

\usepackage{mathtools}
\usepackage{flushend}
\usepackage{float}

\newtheorem{definition}{Definition}

\makeatletter
\def\ps@IEEEtitlepagestyle{%
  \def\@oddfoot{\mycopyrightnotice}%
  \def\@evenfoot{}%
  }

\def\mycopyrightnotice{%
  {\footnotesize 978-1-7281-0858-2/19/\$31.00 © 2019 IEEE \hfill}%
  \gdef\mycopyrightnotice{}%
  }

\begin{document}

\title{Scheduling in the Presence of Data Intensive\\ Compute Jobs
}

\author{\IEEEauthorblockN{Amir Behrouzi-Far and Emina Soljanin}
\IEEEauthorblockA{\textit{Department of Electrical and Computer Engineering, Rutgers University} \\
Piscataway, New Jersey 08854, USA \\
\{amir.behrouzifar,emina.soljanin\}@rutgers.edu}
}

\maketitle

\begin{abstract}
We study the performance of non-adaptive scheduling policies in computing systems with multiple servers. Compute jobs are mostly regular, with modest service requirements. However, there are sporadic data intensive jobs, whose expected  service time is much higher than that of the regular jobs. For this model, we are interested in the effect of scheduling policies on the average time a job spends in the system. To this end, we introduce two performance indicators in a simplified, only-arrival system. We believe that these performance indicators are good predictors of the relative performance of the policies in the queuing system, which is supported by simulations results.

\end{abstract} 
\section{introduction}
Scheduling has been of interest since the emergence of computing systems. Although scheduling problems were studied for a wide  variety of systems, most studies consider compute jobs with deterministic or exponentially distributed or heavy-tail distributed service times. In today's computer systems, a fraction of jobs are data intensive, and their the service time is on average several orders of magnitude longer than that of the regular jobs in the system \cite{chen2010analysis}. Thus previously known scheduling results are not directly applicable in modern environments.

In distributed computing systems, redundancy can reduce fluctuations in the service time of the jobs \cite{joshi2014delay}, and has gained attention both in theoretical works \cite{behrouzi2019redundancy,aktas2019straggler,behrouzi2018effect} and in practical systems \cite{he2010comet,bernardin2006using}. Redundancy is particularly beneficial in systems with high variability in job service times \cite{gardner2017redundancy}.

In a queuing system, submitting redundant copies of an arriving job to several servers has the following benefits. First, by bringing \textit{diversity} to the queuing times, redundancy helps jobs to find a shorter queue than they would find without redundancy. Second, in the event of an unexpected slowdown in a server the redundant copies can be served on (probably) faster servers. Stragglers \cite{ananthanarayanan2013effective}, can significantly increase the average service time and its variability \cite{aktas2019straggler}. By running jobs concurrently at multiple servers, a jobs' service time is the shortest service time experienced by the copies of that job.

Since multiple servers have to be selected for each arriving job the scheduling problem could be exponentially harder in a system with redundancy than in its no-redundancy counterpart. 
The role of redundancy,  which is to \textit{diversify} the queuing/service time, also adds a new dimension to the scheduling problem, and the classical scheduling policies, designed for no-redundancy systems, may not have the expected performance in systems with redundancy.

In this work, we study a queuing system where most of the arriving jobs have (on average and relatively) modest service requirements, and there are sporadic job arrivals whose service are (on average and comparatively) very slow. We assume that jobs get scheduled upon arrival but the service requirement of jobs are not known upon arrival. In order to prevent fast jobs from being penalized in the queues with slow jobs, each job gets redundantly submitted to a few servers upon arrival. Once the first copy of a job started the service the redundant copies get cancelled. This way redundancy brings no extra cost to the system and it only helps a job to find shorter queues. 

Analysis of the average response time of the jobs in queuing systems with redundancy is known to be a hard problem. To develop an insight about the performance of scheduling policies, we studied an \textit{only-arrival} system, where jobs get scheduled into queues redundantly, without departure. The statistics of queues' occupancy in the only-arrival system can then be studied using the classical Urns\&Balls problem. By making this connection, we introduced two performance indicators in the only-arrival system, which we believed are good predictors for the relative performance of the scheduling policies in queuing systems with redundancy. Through simulations, we observed that that really is the case for the considered systems, as the policy which ranks higher by the performance indicators showed better performance.

The scheduling policies that we study are \textit{non-adaptive}, in the sense that for every job arrival the queue selecting process is independent of the state of the queues. We consider random policy, where an arriving job gets scheduled to a number of queues chosen uniformly at random, and round-robin policy, where queues are chosen in a round-robin fashion. Then we propose a scheduling policy based on combinatorial block designs, which according to our performance indicators performs better in a queuing system with redundancy. This claim gets validated by our simulations result.

From the simulations analysis, we also made the following observations. First, the performance of the scheduling policies is fundamentally different in systems with sporadic data intensive jobs, compared to their performance with classical service models. In particular, round-robin policy, which provides good load balancing, performs worse than the random policy. There is a subtle intuition behind this phenomenon. In classical job service time models, the load at each server could be approximately measured by the number of queued jobs. With this measure of load, round-robin policy provides the best average load balancing. However, with sporadic data intensive compute jobs, the number of jobs is not a good measure of the actual load on the servers. For example, a server with only one data intensive job could have more compute load than a server with hundreds of regular jobs. In systems with redundancy, if all the redundant copies of a job get queued behind the copies of a data intensive job then redundancy brings no benefit to the performance. The round-robin policy increases the chance of this phenomenon \cite{behrouzi2019redundancy}.

This observation motivated us to think about new scheduling policies that perform better in the presence of sporadic data intensive jobs. We proposed a scheduling policy, based on combinatorial block designs, which chooses servers for jobs in a way that the number of \textit{overlapping} servers between consequent jobs is minimized. Among all, we focus on Balanced and Incomplete Block Designs (BIBD), and call the related policy \textit{BIBD scheduling} policy.

\section{Problem statement and summary of results}

\begin{figure}[t]
        \centering
        \includegraphics[width=7.8cm]{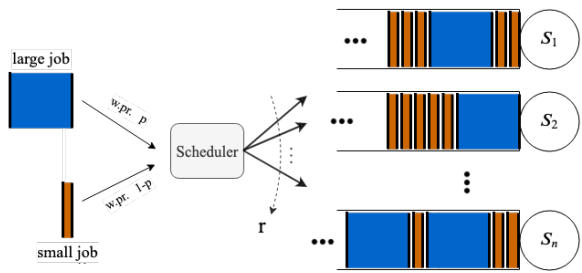}
        \caption{System model.}
        \label{fig:sysModel}
    \end{figure}

We studied the performance of the queuing system shown in Fig.~\ref{fig:sysModel}, with $n$ identical servers. Upon an arrival, $r$ copies of the job get scheduled into $r$ servers. Once the first copy enters service the redundant copies get cancelled. The service time of a copy is sampled from a fast Exponential distribution if the job is a regular one and from a slow Exponential distribution if the job is a data intensive one. Let's define the Exponential random variable $\tau$, with rate $\mu_1$, to be the service time of regular jobs and
    \begin{equation}
        q=\frac{\textup{average service time of data intensive jobs}}{\textup{average service time of regular jobs }}.
        \label{q}
    \end{equation}{}
The average service time of data intensive jobs is the random variable $q\tau$, which is an Exponential with rate $\mu_1/q$. We model the frequency of different types of jobs by $p\in[0,1]$, such that an arriving job is a data intensive one with probability $p$ and is a regular one with probability $1-p$.

For designing a scheduling policy in the presence of data intensive jobs, the following considerations should be taken into account. First, load balancing should not be compromised, since any systematic unbalanced assignment of jobs to the servers can degrade the performance. Load balancing has been the central objective for designing scheduling policies in queuing systems. Nevertheless, with the presence of the data intensive jobs, the ``number" of jobs is not a precise measure of the load on a server. Therefore, the classical load balancing policies could easily fail to evenly distribute the workload among the servers. On the other hand, putting the copies of the two types of jobs into the same queues would penalize the regular jobs by making them wait for the departure of the data intensive jobs. If the service requirement of jobs were known upon arrival the optimal scheduling policy would be to divide servers to two groups and dedicate each group exclusively to one type of jobs. However, since the service requirement of the jobs are not known at the time of scheduling, the best that it can be done is to minimize the chance of the regular jobs to overlap with the data intensive jobs. Certainly, the introduction of redundancy can reduce the probability of overlaps. Nevertheless, using techniques form combinatorial block design, we show that it is possible to further reduce the chance of overlaps, with scheduling policies based on Balanced and Complete Block Designs (BIBD). To prove our arguments, in the following, we develop an analogy to the classical Urns\&Balls problem and drive our performance indicators in the analogue problem. Interested readers may refer to \cite{behrouzi2019redundancy} for detailed analysis.

\subsection{An Urns and Balls Analogy}
Let's consider the only-arrival queuing system. In the analog Urns\&Balls problem, the urns are the queues and the balls are the jobs. Let's define $N_i^T$, $i=1,2,\dots,n$, as the random variable for the number of balls in urn $i$ after throwing $T$ balls redundantly into $n$ urns. Let's also define the random variable $X$, as the number of overlaps in the set of chosen servers for consequent jobs. Based on the defined random variables we introduce the following performance indicators.

    \begin{figure}[t]
        \centering
        \includegraphics[width=\columnwidth]{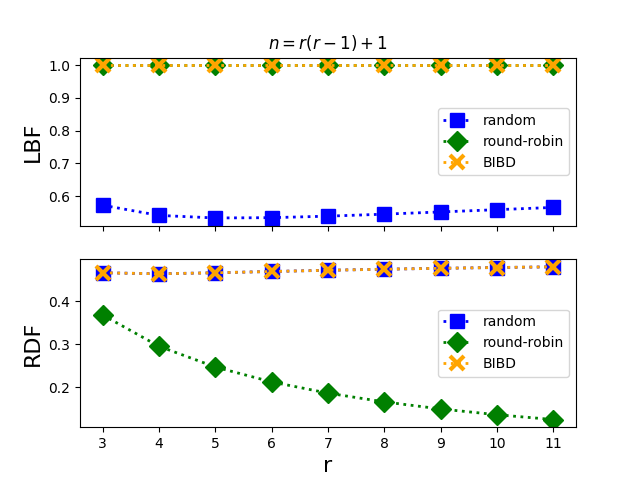}
        \caption{LBF and RDF for non-adaptive scheduling policies as a function of redundancy level $r$ and number of urns (servers) being $n=r(r-1)+1$. }
        \label{fig:lbfrdf}
    \end{figure}
    
    \begin{figure*}[t!]
        \centering
        \includegraphics[width=11cm,trim={1.8cm 0 2.7cm 0},clip]{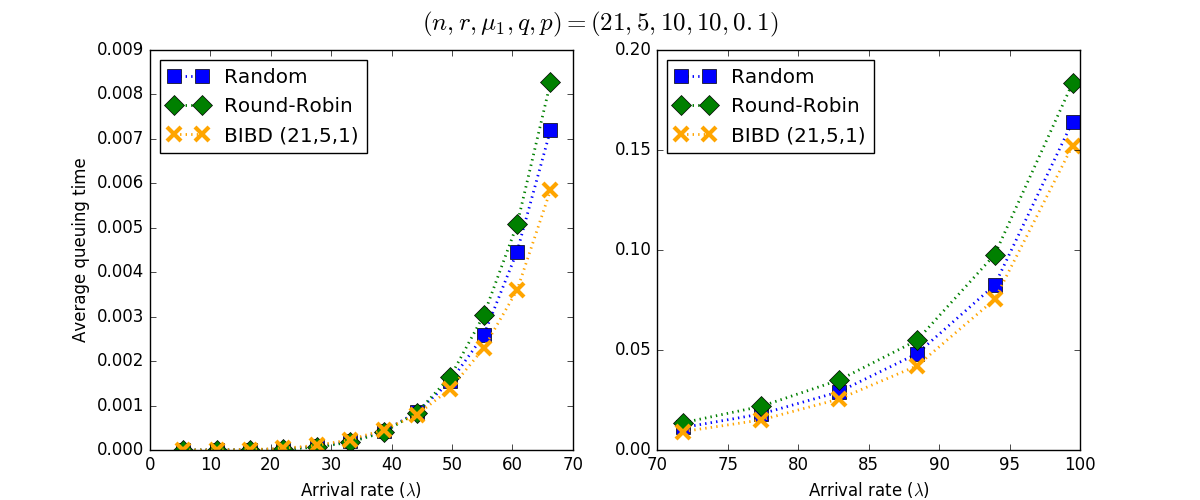}
        \caption{Average queuing time for non-adaptive scheduling policies, with $(n,r,\mu_1,q,p)=(21,5,10,10,0.1)$. }
        \label{fig:21_5_10_10}
    \end{figure*}
    
\begin{definition}
The Load Balancing Factor (LBF) and the Redundancy Diversity Factor (RDF) of a (scheduling) policy are defined as,
    \begin{equation}
        \textup{LBF}_{policy}\coloneqq\frac{\mathbb{E}\left[N_{1:n}^T\right]}{\mathbb{E}\left[N_{n:n}^T\right]},
    \label{def:LBF}
    \end{equation}{}
    \begin{equation}
        \textup{RDF}_{policy}\coloneqq\frac{1}{\mathbb{E}[X^2]}.\quad
    \label{def:RDF}
    \end{equation}
\end{definition}

The behaviour of the performance metrics are plotted in Fig. \ref{fig:lbfrdf}. BIBD scheduling policy achieves the highest of both indicators, while the round-robin policy fails at RDF and the random policy fails at LBF. 

The reasons we believe that LBF and RDF are good performance indicators are as follows. First, LBF can predict that how well a scheduling policy can (on average) distribute jobs among servers. Note that, since the service requirement of jobs are not know upon arrival the best a scheduling policy can do, in terms of load balancing, is to assign equal number of jobs to every servers. However, since there are data intensive jobs, we argue that LBF is not the only performance indicator, which brings us to the second reason. RDF measures the variance of the number of overlapping servers between consequent jobs, for a given arrival process. This is important in the presence of data intensive jobs, since high variance of the overlapping servers essentially means that there is a high chance that some subsequent jobs will overlap in many servers, out of the ones they have been submitted to. This phenomenon penalizes regular jobs, by submitting their replicas into the servers that are already occupied by the replicas of data intensive jobs. On the other hand, with small variance, the number of overlapping servers is moderate between consequent jobs, which in turn increases the chance of a regular job to find shorter queues.

\subsection{Simulation Results}
Fig. \ref{fig:21_5_10_10} shows the average queuing time of the jobs versus the arrival rate, for three scheduling policies; random, round-robin and BIBD. The results are for a queuing system with $n=21$ queues, where each job is submitted to $r=5$ queues, the service time of regular jobs and data intensive jobs are Exponentially distributed with rate $\mu_1=10$ and $10\mu_1=100$, respectively. An arrival is a data intensive job with probability $0.1$ and is a regular job with probability $0.9$.  The left and the right sub-figures show the result for the low to medium and medium to high arrival rates, respectively. In both regions, BIBD scheduling policy out performs the random policy, by $10\%$, and the round-robin policy, by $20\%$. This is inline with what the proposed performance indicator suggests in Fig. \ref{fig:lbfrdf}. According to this figure, BIBD policy provides the highest load balancing and diversity in redundancy at the same time. Further, the round-robin policy is inferior compared to random policy, which is because of the presence of data intensive jobs.

\section*{Acknowledgement}
Part of this research is based upon work supported by the NSF grants No.\ CIF-1717314 and CCF-1559855.

\bibliographystyle{IEEEtran}
\bibliography{ref}

\begin{thebibliography}{1}
\providecommand{\url}[1]{#1}
\csname url@samestyle\endcsname
\providecommand{\newblock}{\relax}
\providecommand{\bibinfo}[2]{#2}
\providecommand{\BIBentrySTDinterwordspacing}{\spaceskip=0pt\relax}
\providecommand{\BIBentryALTinterwordstretchfactor}{4}
\providecommand{\BIBentryALTinterwordspacing}{\spaceskip=\fontdimen2\font plus
\BIBentryALTinterwordstretchfactor\fontdimen3\font minus
  \fontdimen4\font\relax}
\providecommand{\BIBforeignlanguage}[2]{{%
\expandafter\ifx\csname l@#1\endcsname\relax
\typeout{** WARNING: IEEEtran.bst: No hyphenation pattern has been}%
\typeout{** loaded for the language `#1'. Using the pattern for}%
\typeout{** the default language instead.}%
\else
\language=\csname l@#1\endcsname
\fi
#2}}
\providecommand{\BIBdecl}{\relax}
\BIBdecl

\bibitem{chen2010analysis}
Y.~Chen, A.~S. Ganapathi, R.~Griffith, and R.~H. Katz, ``Analysis and lessons
  from a publicly available google cluster trace,'' \emph{EECS Department,
  University of California, Berkeley, Tech. Rep. UCB/EECS-2010-95 94}, 2010.

\bibitem{joshi2014delay}
G.~Joshi, Y.~Liu, and E.~Soljanin, ``On the delay-storage trade-off in content
  download from coded distributed storage systems,'' \emph{IEEE Journal on
  Selected Areas in Communications}, vol.~32, no.~5, pp. 989--997, 2014.

\bibitem{behrouzi2019redundancy}
A.~Behrouzi-Far and E.~Soljanin, ``Redundancy scheduling in systems with
  bi-modal job service time distribution,'' \emph{arXiv preprint
  arXiv:1908.02415}, 2019.

\bibitem{aktas2019straggler}
M.~F. Aktas and E.~Soljanin, ``Straggler mitigation at scale,'' \emph{arXiv
  preprint arXiv:1906.10664}, 2019.

\bibitem{behrouzi2018effect}
A.~Behrouzi-Far and E.~Soljanin, ``On the effect of task-to-worker assignment
  in distributed computing systems with stragglers,'' in \emph{2018 56th Annual
  Allerton Conference on Communication, Control, and Computing
  (Allerton)}.\hskip 1em plus 0.5em minus 0.4em\relax IEEE, 2018, pp. 560--566.

\bibitem{he2010comet}
B.~He, M.~Yang, Z.~Guo, R.~Chen, B.~Su, W.~Lin, and L.~Zhou, ``Comet: batched
  stream processing for data intensive distributed computing,'' in
  \emph{Proceedings of the 1st ACM symposium on Cloud computing}.\hskip 1em
  plus 0.5em minus 0.4em\relax ACM, 2010, pp. 63--74.

\bibitem{bernardin2006using}
J.~Bernardin, P.~Lee, and J.~Lewis, ``Using execution statistics to select
  tasks for redundant assignment in a distributed computing platform,'' Aug.~15
  2006, uS Patent 7,093,004.

\bibitem{gardner2017redundancy}
K.~Gardner, M.~Harchol-Balter, A.~Scheller-Wolf, M.~Velednitsky, and
  S.~Zbarsky, ``Redundancy-d: The power of d choices for redundancy,''
  \emph{Operations Research}, vol.~65, no.~4, pp. 1078--1094, 2017.

\bibitem{ananthanarayanan2013effective}
G.~Ananthanarayanan, A.~Ghodsi, S.~Shenker, and I.~Stoica, ``Effective
  straggler mitigation: Attack of the clones,'' in \emph{Presented as part of
  the 10th $\{$USENIX$\}$ Symposium on Networked Systems Design and
  Implementation ($\{$NSDI$\}$ 13)}, 2013, pp. 185--198.

\end{thebibliography}
\end{document}